\documentclass[sn-mathphys]{sn-jnl}

\jyear{2021}%

\theoremstyle{thmstyleone}%
\theoremstyle{thmstyletwo}%

\theoremstyle{thmstylethree}%

\begin{document}

\title[]{Multimodal Remote Sensing Image Registration Based on Adaptive Multi-scale PIIFD}

\author[1]{\fnm{Ning} \sur{Li}}\email{lnmmdsy@bupt.edu.cn}\equalcont{These authors contributed equally to this work.}

\author*[2]{\fnm{Yuxuan} \sur{Li}}\email{li.yuxuan@bupt.edu.cn}
\equalcont{These authors contributed equally to this work.}

\author[3]{\fnm{Jichao} \sur{jiao}}\email{jiaojichao@bupt.edu.cn}

\affil*[1]{\orgdiv{School of Electronic Engineering}, \orgname{Beijing University of Posts and Telecommunications}, \orgaddress{\city{Beijing}, \ \country{People's Republic of China}}}


\abstract{In recent years, due to the wide application of multi-sensor vision systems, multimodal image acquisition technology has continued to develop, and the registration problem based on multimodal images has gradually emerged. Most of the existing multimodal image registration methods are only suitable for two modalities, and cannot uniformly register multiple modal image data. Therefore, this paper proposes a multimodal remote sensing image registration method based on adaptive multi-scale PIIFD(AM-PIIFD). This method extracts KAZE features, which can effectively retain edge feature information while filtering noise. Then adaptive multi-scale PIIFD is calculated for matching. Finally, the mismatch is removed through the consistency of the feature main direction, and the image alignment transformation is realized. The qualitative and quantitative comparisons with other three advanced methods shows that our method can achieve excellent performance in multimodal remote sensing image registration.}

\keywords{image registration, remote sensing, multi-scale, multimodal, PIIFD}



\maketitle

\section{Introduction}\label{sec1}

Image registration is the process of overlaying two or more images of the same scene taken at different times, from different viewpoints, and/or by different sensors. It geometrically aligns two images—the reference and sensed images\cite{zitova2003image}. Multi-sensor data usually provides supplementary information about the area being measured to obtain an image with more information. The main problems to be solved for registration of multimodal remote sensing images are: the difference in image intensity and scale caused by different sensors, which may make the local description of the corresponding feature points different, or the corresponding feature points do not exist, which will cause incorrect matching and image registration. As most methods cannot register multiple modal images at the same time, the existing registration methods can be roughly divided into two categories: area-based registration method and feature-based registration method \cite{tondewad2020remote}.

The area-based registration method mainly uses image grayscale information to establish a similarity measure for image registration, and the most representative method is mutual information (MI) \cite{wells1996multi,viola1997alignment}.In addition, there are Cross Correlation (CC)\cite{goshtasby1986region}, phase correlation based on Fast Fourier Transform, etc. However, the existing area-based methods have different degree of problems in image modality, intensity transformation, complex spatial transformation, computational complexity, and so on. Therefore, they are not suitable for multimodal remote sensing image registration.

Compared with area-based methods, feature-based methods are more robust in dealing with problems such as image intensity changes and image noise. Feature-based multimodal remote sensing image registration mainly consists of three steps, feature extraction, feature description, and feature matching. Feature extraction mainly extracts the significant structure in the image. Because of the stability of point features, most methods mainly extract point features. Scale Invariant Feature Transform (SIFT)\cite{lowe2004sift} is a classical point feature with good scale and rotation invariance, so SIFT and its improved methods are widely used in multimodal image registration. For example, Sedaghat et al.\cite{sedaghat2011uniform} improved the SIFT algorithm in the feature selection strategy, called uniform robust SIFT (UR-SIFT). KAZE feature\cite{alcantarilla2012kaze} is a multi-scale two-dimensional feature detection and description algorithm in non-linear scale space. Pourfard et al.\cite{pourfard2021kaze} use KAZE to extract features to reduce speckle noise of SAR images and register them. Feature description refers to the generation of specific descriptors for extracted feature points in preparation for subsequent feature matching. Based on the SIFT algorithm, Ma et al.\cite{ma2016remote} introduced a new gradient definition and enhanced feature matching method (PSO-SIFT) to overcome image intensity differences between remote image pairs. Chen et al.\cite{chen2010partial} proposed a Partial Intensity Invariant Feature Descriptor (PIIFD) for multi-source retinal image registration, which is intensity and rotation invariant, but cannot handle scale differences, the original PIIFD has no scale-invariant. Radiation-variation insensitive feature transform (RIFT)\cite{li2019rift} proposed a maximum index map (MIM) for feature description, and they used phase consistency to compute the MIM for feature description. It has rotation-invariance, but no scale-invariance. DU et al.\cite{du2018infrared} proposed a scale-invariant PIIFD (SI-PIIFD) feature and a robust feature matching method, which by calculating several fixed-range feature description regions for each feature point to achieve multi-scale PIIFD. Gao et al.\cite{gao2021multi} proposed a multi-scale Harris-PIIFD image registration algorithm framework, which calculates PIIFD descriptors by constructing scale space at the extracted Harris features. This reduces the impact of scale differences in multimodal images. However, both multi-scale PIIFD algorithms need to compute multiple PIIFD at the features, which increases the computational complexity.

Our method has a few innovations as follows. 1) An adaptive multi-scale PIIFD (AM-PIIFD) is proposed to exclude the nonlinear intensity differences of different modal images, accurately determine the description position and reduce the computational complexity. 2) For the characteristics of remote sensing images, the elimination of mismatching is performed by using the main direction consistency, which improves the accuracy of image matching. The experimental results in the collected public data show that the method has excellent and stable performance in multimodal remote sensing image registration. It has good generality and strong practical application value. This article is organized as follows. Section 2 of the article introduces the KAZE algorithm, the improved PIIFD feature descriptor, and the mismatch elimination method, and the experimental results are presented in Section 3. Section 4 summarizes and describes the future work.

\section{Proposed method}\label{sec2}

\subsection{Feature Extraction}
Edge details are more important in multi-modal registration methods based on feature extraction\cite{wang2020feature}. In linear filtering methods, such as Gaussian scale space, the details and noise are smoothed to the same degree, resulting in blurred boundaries and reduced details. But the nonlinear diffusion filtering used by the KAZE algorithm can solve the related problems well. The nonlinear diffusion filtering method is usually described by a nonlinear partial differential equation, as in equation (1)

\begin{equation}
\label{eq:anis}
\frac{\mathrm d L}{\mathrm d t} = div\left( c(x,y,t). \nabla L \right)
\end{equation}

\begin{equation}\label{eq:conduc}
c(x,y,t) = g(\lvert\nabla L_{\sigma}(x,y,t)\rvert)
\end{equation}

Where$ L $is the brightness of the image, time $ t $ is the scale parameter, $ div $ and $ \nabla $ denote the gradient and scatter, respectively, $ c $ is the conductivity function, and $(x,y)$ is the pixel coordinate of the image. Perona and Malik \cite{perona1990scale} proposed to let the function c depend on the gradient magnitude, as in equation (2), The gradient magnitude of the image controls the diffusion of the different scale levels so that it has larger values at the background regions of the image and smaller values at the edges, making it smooth in the regions without crossing the edges and preventing the edges from being smoothed.

where $\nabla L_{\sigma}$ is the gradient of the Gaussian smoothed version (g) of the original image I. Weickert \cite{weickert2001efficient} proposed a diffusion function in which the smoothing on both sides of the edge is much stronger than the smoothing across the edge, as in equation (3). K is a contrast factor controlling the level of diffusion, which can determine how much edge information is retained. The larger the value, the less edge information is retained. By using the g3 equation, the blurred region, retains the sharpedge that we are concerned about.

\begin{equation}\label{eq:g3}
g_3 = ~\left\{ \begin{matrix}
{~~~~~~~~~~~~~~~~1~~~~~~~~~~~~~~~,~\lvert {\nabla L}_{\sigma} \rvert^{2}~ = ~0} \\
{1 - ~{\exp\left( {- ~\frac{3.315}{\left( \frac{\lvert {\nabla L}_{\sigma} \rvert}{k} \right)^{8}}} \right)}~~~~~~,~\lvert {\nabla L}_{\sigma} \rvert^{2} ~> ~0~~~} \\
\end{matrix} \right.
\end{equation}

Next, compute the nonlinear scale space. The scale space is discretized and then arranged in logarithmic steps in a series of $O$ octaves and $S$ sub-levels, the relationship between the layers is as follows.

\begin{equation}\label{eq:sigma}
\sigma_{i}\left( o_{i},{~s}_{i} \right)~ = ~\sigma_{0}2^{o_{i} + s_{i}/S}
\end{equation}

where $\sigma_0$ is the benchmark scale level,$\nonumber o \in [0...O-1], s \in [0...S-1], \nonumber i\in[0...N]$, and $ N $ is the total number of filtered images. Then, the discrete scale level of pixel unit $\sigma_i$ in equation (4) needs to be converted to time unit, since nonlinear diffusion filtering is defined in time. In the case of Gaussian scale space, convolution of the image using a Gaussian kernel with standard deviation $\sigma$ is equivalent to filtering the image with duration $t$ = $\frac{\sigma^2}{2}$, and we apply this transformation to convert the scale space $\sigma_i$ to the evolution time ti. According to a set of evolution time t, all images in the nonlinear scale space can be obtained by using the AOS\cite{alcantarilla2012kaze} algorithm. Finally, the feature points are obtained by finding the local maxima of the normalized Hessian determinant at different scales.

\subsection{Adaptive multi-scale PIIFD}	
After extracting the local features, we need to describe the local information around the feature points and generate descriptors to facilitate matching.
\subsubsection{Calculate feature region}
In multimodal remote sensing images, different modal images generally have different resolutions and different visual areas, which leads to the change of scale. PIIFD uses a fixed neighborhood size (generally 40$\times$40) and cannot show the variation of feature scale. In SIFT algorithm, it detects feature points in scale space and provides its scale information for each feature point. Then, the neighborhood size of the extracted descriptor is determined based on the scale information to achieve scale invariance. Therefore, learning from the SIFT algorithm ,we need adaptive neighborhood regions to achieve scale invariance and thus describe the features accurately.

\begin{equation}\label{eq:mu}
\mu_{i} = offset*2^{o_{i} + {({s_{i} + ~\lambda_{i}})}/S}
\end{equation}

\begin{equation}\label{eq:xyz}
x~ = {~(x,y,\lambda)}^{T}
\end{equation}

\begin{equation}\label{eq:x}
\hat{x}~= (\frac{\partial^{2}L}{\partial x^{2}} )^{- 1}\frac{\partial L}{\partial x}
\end{equation}

Since the interest point scale of each response is different, while detecting the feature point response, it is necessary to obtain the scale factor of each interest point to calculate the feature region. We set the scale factor of the characteristic points of the response as $\mu$, as shown in equation (5).
Where  $offset$  is a constant, usually we take 1.6. $o_i$ and $s_i$ is the octave index and sub-level index to which the current feature point belongs, such as equation (4). $\lambda_i$ is a variable, calculated by calculating the subpixel approximate coordinates of the feature points, such as equation (6,7), Where $L(x)$ is an approximation of the Laplacian operator, and X is the approximation of spatial coordinates, it can be found $\nonumber \lambda \in \lbrack - 1~,~1\rbrack$. By equation (5), Scale Factor $\mu_i$ is determined by the scale space in which the feature is located. The larger the scale factor, the larger the current feature response region. Due to the need to handle the inverse gradient problem, the detection region is set to square , and the neighborhood size of our AM-PIIFD is set to 
$\nonumber ~(k\mu)~\times~(k\mu)$. The default value of $k$ is 6, which is measured by experiments. This takes a variable detection range for each feature. An adaptive neighborhood determined by the size of the feature points is used to achieve scale invariance.

\subsubsection{Extract Descriptor}
After the feature description area is determined, the descriptor is extracted. First, the magnitude and direction of the image gradient are calculated, and a continuous average square gradient is used to solve the opposite gradient problem, so as to obtain better accuracy and computational efficiency. The extraction area is composed of 16 small squares, and the area of each small square is $\nonumber\left( \frac{k\mu}{4} \right)^{2}$, corresponding to a directional histogram By calculating the sum of opposite directions, the direction histogram with 16 bins uniformly covering $\left. 0 \right.\sim 2\pi~\left( 0{^\circ},~22.5{^\circ},.~.~.,~337.5{^\circ} \right)$ is converted into a degenerate direction histogram with only 8 bins uniformly covering $\left. 0 \right.\sim\pi\left( 0{^\circ},~22.5{^\circ},.~.~.,~157.5{^\circ} \right)$, as shown in Fig. 1. This enables invariance when the gradient orientation rotates by $180{^\circ}$ . In order to solve the opposite gradient problem, PIIFD uses the linear combination of two sub descriptors. For example, the original direction histogram matrix H and its rotation version $Q = rot\left( H,180^{◦} \right)$, for example, the square where $H$ is 4 * 4 is defined as:
\begin{equation}
H = ~\begin{bmatrix}
{H_{11},H_{12},H_{13},H_{14}} \\
{H_{21},H_{22},H_{23},H_{24}} \\
{H_{31},H_{32},H_{33},H_{34}} \\
{H_{41},H_{42},H_{43},H_{44}} \\
\end{bmatrix}
\end{equation}
Set $H_I$ and $Q_I$ is row i of H and Q respectively, so PIIFD is calculated as:
\begin{equation}
H = ~\begin{bmatrix}
\left( H_{1} + ~rot\left( H_{1},180^{\circ} \right) \right) \\
\left( H_{2} + ~rot\left( H_{2},180^{\circ} \right) \right) \\
{C\lvert \left( H_{3} - ~rot\left( H_{3},180^{\circ} \right) \right) \rvert} \\
{C\lvert \left( H_{4} - ~rot\left( H_{4},180^{\circ} \right) \right) \rvert} \\
\end{bmatrix}
\end{equation}

Where C is a parameter that adjusts the scale of the local descriptor. Formula above gives that PIIFD is a 4*4*8 matrix. The 128-dimensional descriptor vector is generated and normalized to unit length for feature point description.

\begin{figure}[htbp]
\centering
\includegraphics[width=0.3\textwidth]{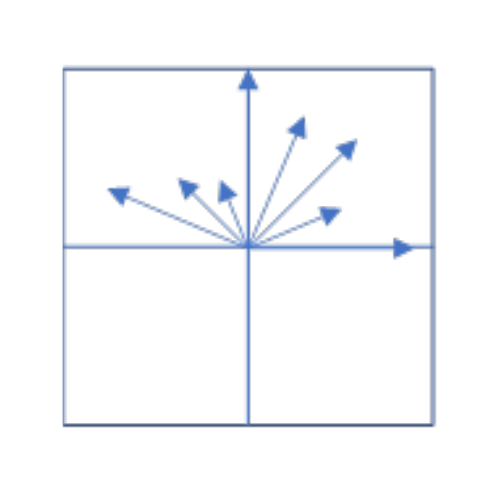}
\caption{ degraded orientation histogram }
\label{fig:figure1}
\end{figure}

\subsection{Feature matching}
Firstly, BBF (best-bin-First)\cite{beis1997shape} matching method is created by using the generated feature descriptors, and bilateral matching is performed to obtain the initial matching of feature points. Then use the consistency of the main orientation to remove the mismatch. The main orientation of the feature point ensures the rotation invariance of the feature point. When the image is rotated to the same position, theoretically the main orientation of the features is the same between correctly matched pairs, while the orientation of the incorrect matches is different. Therefore, when the image is rotated to the same position, the main orientation angle difference between the two points is less than 5°, which is regarded as a correct matching pair. The specific method is as follows: the main orientation of the initial matching is$
{~\varnothing}_{x} = \left\{ {\varnothing x}_{i} \right\}_{i = 1}^{N}$ and $\varnothing_{y} = \left\{ {\varnothing y}_{j} \right\}_{j = 1}^{N}$, $N$ is the number of matching pairs, $\varnothing$ is the main orientation angle of multimodal image feature points, $x$,$y$ is the image modality, In this paper, $\in ~\lbrack 1\rbrack,~y~ \in ~\left\lbrack 1~,2,...~,7 \right\rbrack$, Therefore, the main orientation angle difference $
\mathrm{\Delta}\varnothing$ is:

\begin{equation}
\mathrm{\Delta}\varnothing = \left\{ {\mathrm{\Delta}\varnothing}_{i}~ \vert ~{\mathrm{\Delta}\varnothing}_{i} = {\varnothing x}_{i} - {\varnothing y}_{i} \right\}_{i = 1}^{N}
\end{equation}

Due to the error, we use the histogram for statistics 
$\mathrm{\Delta}\varnothing$, with 5° as the interval, the range of the histogram's x-axis is [0°, 360°), and the y-axis counts the number of 
$\mathrm{\Delta}\varnothing$ included in the corresponding interval. We take the feature pair that contains the most feature pairs in the histogram as our correctly matched feature point pair. Finally, RANSAC\cite{fischler1981random} is used to further remove the matching error and get the matching result. Finally, according to the matching pair results, one of similarity transformation, affine transformation and projection transformation is used to estimate the parameters and transform the model, and the least squares method is used to calculate the model parameters.

\section{Our experiments and result}\label{sec3}

\begin{figure*}[htbp]
\setlength{\abovecaptionskip}{-0.5cm}   
\begin{center}
   \includegraphics[width=1.0\linewidth]{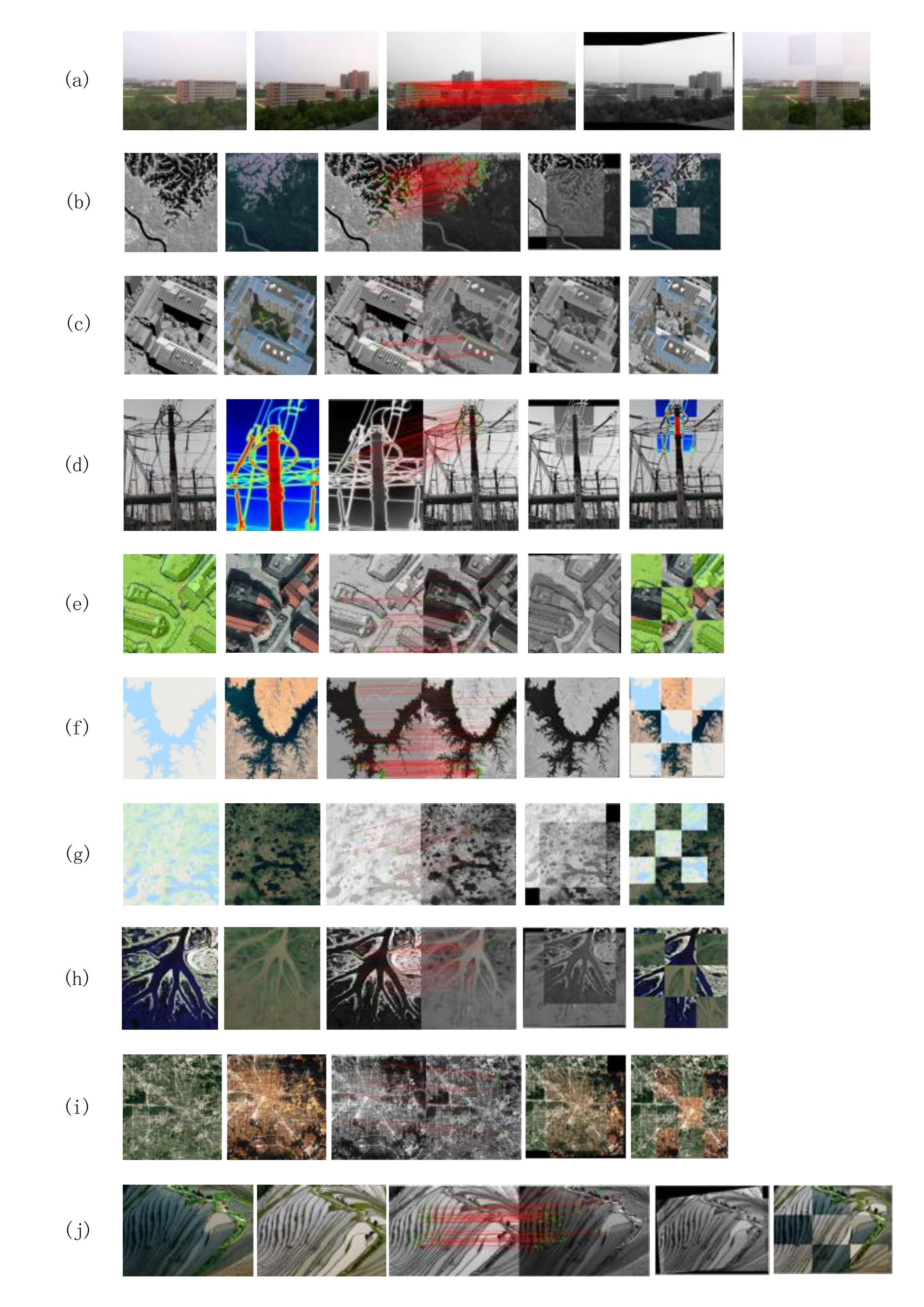}
\end{center}
   \caption{Ten pairs of image registration results, which are composed of five parts: reference image, sensed image, image matching result, gray mosaic image, RGB mosaic image. The images modal are:(a) optical-optical, (b) infrared-optical, (c) infrared-optical, (d) depth-optical, (e) depth-optical, (f) map-optical, (g) map-optical, (h) SAR-optical, (i) day-night, (j) CrossSeason.}
\label{fig:long}
\label{fig:onecol}
\end{figure*}

\begin{sidewaystable}[htbp]
\sidewaystablefn%
\begin{center}
\begin{minipage}{\textheight}
\caption{Correct matching rate(CMR) comparison by four registration methods}\label{table1}
\begin{tabular*}{\textheight}{@{\extracolsep{\fill}}ll|cccccccccc@{\extracolsep{\fill}}}
\hline
\multirow{2}{*}{Method} & \multirow{2}{*}{Index} & \multicolumn{10}{c}{image group} \\ \
 &  & a & b & c & d & e & f & g & h & i & j \\
\hline
\multirow{3}{*}{PSO-SIFT} & Nc & 76 & 84 & 2 & 2 & 0 & 36 & 9 & 3 & 13 & 32 \\
 & N & 488 & 278 & 282 & 94 & 152 & 288 & 124 & 228 & 114 & 187 \\
 & CMR & 0.156 & 0.302 & - & - & - & 0.125 & 0.072 & - & 0.114 & 0.171 \\
 \hline
\multirow{3}{*}{SURF-PIIFD-RPM} & Nc & 98 & 8 & 3 & 15 & 8 & 59 & 13 & 3 & 14 & 26 \\
 & N & 278 & 53 & 105 & 114 & 101 & 104 & 78 & 103 & 85 & 80 \\
 & CMR & 0.352 & 0.151 & - & 0.131 & 0.079 & 0.567 & 0.167 & - & 0.164 & 0.325 \\
 \hline
\multirow{3}{*}{RIFT} & Nc & 99 & 115 & 5 & 31 & 57 & 113 & 69 & 38 & 60 & 39 \\
 & N & 439 & 359 & 285 & 290 & 302 & 394 & 285 & 324 & 349 & 243 \\
 & CMR & 0.226 & 0.32 & - & 0.106 & 0.189 & 0.287 & 0.242 & 0.117 & 0.172 & 0.16 \\
\hline
\multirow{3}{*}{Proposed} & Nc & 346 & 61 & 45 & 46 & 26 & 103 & 14 & 17 & 17 & 33 \\
 & N & 423 & 103 & 154 & 153 & 112 & 151 & 30 & 61 & 57 & 80 \\
 & CMR & \textbf{0.818} & \textbf{0.592} & \textbf{0.292} & \textbf{0.301} & \textbf{0.232} & \textbf{0.682} & \textbf{0.467} & \textbf{0.279} & \textbf{0.298} & \textbf{0.413}\\  
\hline
\end{tabular*}
\end{minipage}
\end{center}
\end{sidewaystable}

\begin{sidewaystable}[htbp]
\sidewaystablefn%
\begin{center}
\begin{minipage}{\textheight}
\caption{Root mean square error (RMSE) comparison by four registration methods}\label{table2}
\begin{tabular*}{\textheight}{@{\extracolsep{\fill}}ll|cccccccccc@{\extracolsep{\fill}}}
\hline
\multirow{2}{*}{Method} & \multirow{2}{*}{Index} & \multicolumn{10}{c}{image group} \\
 &  & a & b & c & d & e & f & g & h & i & j \\
\hline
PSO-SIFT & RMSE & 1.6191 & 1.678 & - & - & - & 3.6323 & 2.564 & - & 3.6235 & 3.6572 \\
SURF-PIIFD-RPM & RMSE & 4.9091 & 8.8282 & - & 3.1987 & 2.1743 & 3.5364 & 1.8535 & - & 5.1656 & 7.2127 \\
RIFT & RMSE & 1.4256 & 1.2218 & - & 2.572 & \textbf{1.1811} & 3.1811 & 1.7114 & 1.1917 & 2.9835 & \textbf{2.2392} \\
Proposed & RMSE & \textbf{1.0607} & \textbf{1.129} & \textbf{4.1095} & \textbf{2.5679} & 2.2995 & \textbf{3.0109} & \textbf{1.6961} & {1.1365} & \textbf{2.3956} & 2.9335\\
\hline
\end{tabular*}
\end{minipage}
\end{center}
\end{sidewaystable}

The proposed method is compared with PSO-SIFT\cite{ma2016remote}, SURF-PIIFD-RPM\cite{wang2015robust}, and RIFT\cite{li2019rift}. All experiments are compiled using MATLAB on a laptop with 2.6GHz Intel CPU and 16GB RAM.

\subsection{Data and evaluation}\label{subsec2}
The source of the test images is a public dataset\cite{jiang2021review,yao2022multi,li2019rift,jiang2020contour}.According to the imaging type, it mainly includes visible and visible images, visible and infrared images, visible and depth map, visible and artificially produced rasterized map images, day and night images, seasonal change images, and SAR images and visible images, We select one or two pairs of images from each of these seven types and display them in ten groups a-j. They mainly include the problems of multimodal remote sensing images, such as intensity difference, spatial distortion, rotation, scale difference, and detail difference , noise, etc. Among them, there are obvious differences in scale, intensity, and angle in group (c) images.
The registration results were evaluated using correct matching rate (CMR) and root mean square error (RMSE). The formula for CMR is:

\begin{equation}\label{eq:delta}
CMR = N_{c}/N
\end{equation}
Among them, $N_c$ is the number of correct matching points, and N is the number of established matching points. The larger the CMR, the more accurate the matching, the larger the $N_c$, the more the number of matching pairs. RMSE is an important criterion for image registration quality evaluation, and the formula is:
\begin{equation}\label{eq:rmse}
RMSE = ~\sqrt{\frac{\left( {\sum\limits_{I = 1}^{N}\left\lbrack \left( {x_{ref}^{i} - x_{sen}^{i}} \right)^{2} + \left( {y_{ref}^{i} - y_{sen}^{i}} \right)^{2} \right.} \right\rbrack)}{N}}
\end{equation}
where N is the number of matched feature points, $
\left( x_{ref}^{i},y_{ref}^{i} \right)$ is the position of the i feature point in the reference image,$\left( x_{sen}^{i},y_{sen}^{i} \right)$ is the $i_th$ feature of the registered image point location. In the RMSE metric, the smaller the RMSE, the smaller the difference after image registration.

\subsection{Results and comparison}

The experimental registration results are shown in Fig. 2. The results show that all experimental groups have no obvious deviation and deformation, and can align the images well.

The method performance comparison is shown in Table 1, Table 2, Table 1 shows the CMR results, where the - symbol indicates that the method could not be correctly aligned on that image pair. The reason maybe that the matching pairs are wrong, their CMR is meaningless, so it is not calculated. Our method is better than the other three in terms of matching accuracy and quantity. For group c, the other three registration methods are not scale-invariant and thus cannot match images correctly. Group d,e is challenging for PSO-SIFT due to the principle of depth map imaging, with less texture information and large intensity difference. For group h, the speckle noise unique to SAR images hinders the extraction of features, which also leads to their matching errors. Although our method is ahead of the other three methods on CMR. We are less than the RIFT method in the number of correct matches Nc for some images. The analysis is that the phase coherence is better and easier than KAZE in these images Feature points are detected, which results in the advantage of a higher number of correct matches for RIFT. But in our method CMR accuracy is higher, which is undeniable. In Table 2 RMSE metrics, the meaning of the - symbol is the same as in Table 1, and our method maintains the advantage in most test groups. This shows that our method can register multi-source images more accurately.

\section{Conclusions}

In this paper, we propose a remote sensing image registration. We improved PIIFD has scale characteristics and reduces the multi-scale computational complexity. According to the consistency of the main direction of the feature, the mismatch is distinguished, and the image is accurately aligned. Through experimental analysis, our method has certain advantages in the number of matches and the accuracy of registration, and can well register multimodal remote sensing images. By testing in different scenarios, the method has good robustness and accuracy. In the future, we will collect more kinds of multimodal images for method test application and improve the method.

\section{Declarations}
Conflict of Interest: The authors declare that they have no conflict of interest.

\section{Data availability}

The data that support the findings of this study are available from [\cite{jiang2021review,yao2022multi,li2019rift,jiang2020contour}] but restrictions apply to the availability of these data, which were used under license for the current study, and so are not publicly available. Data are however available from the authors upon reasonable request and with permission of [\cite{jiang2021review,yao2022multi,li2019rift,jiang2020contour}].

\bibliography{sample}


\end{document}